\documentclass[aps,prc,reprint,groupedaddress,twocolumn,superscriptaddress]{revtex4-2}

\usepackage{amssymb}
\usepackage{amsmath}
\usepackage{graphicx}
\usepackage{color}
\usepackage{braket}

\begin{document}


\title{Model-independent determination of the dipole response of $^{66}$Zn using quasi-monoenergetic and linearly-polarized photon beams}


\author{D. Savran}
\email{d.savran@gsi.de}
\affiliation{GSI Helmholtzzentrum für Schwerionenforschung GmbH, 64291 Darmstadt, Germany}

\author{J. Isaak}
\affiliation{Technische Universit\"at Darmstadt, Department of Physics, Institute for Nuclear Physics,
            64289 Darmstadt, Germany}

\author{R.~Schwengner} 
\affiliation{Helmholtz-Zentrum Dresden-Rossendorf, 01328 Dresden, Germany}
\author{R.~Massarczyk}
\affiliation{Los Alamos National Laboratory, Los Alamos, New Mexico 87545, USA}
\author{M.~Scheck}
\affiliation{CEPS, University of the West of Scotland, Paisley PA1 2BE,
             United Kingdom}
\affiliation{SUPA, Scottish Universities Physics Alliance, United Kingdom}
\author{W.~Tornow}
\affiliation{Department of Physics, Duke University, Durham,
             North Carolina 27708, USA}
\affiliation{Triangle Universities Nuclear Laboratory, Durham, 
             North Carolina 27708, USA}
\author{G.~Battaglia}
\affiliation{University of Strathclyde, Glasgow G4 0NG, United Kingdom}
\author{T.~Beck}
\affiliation{Facility for Rare Isotope Beams, Michigan State University, East Lansing, Michigan 48824, USA}
\author{S.~W.~Finch}
\affiliation{Department of Physics, Duke University, Durham,
             North Carolina 27708, USA}
\affiliation{Triangle Universities Nuclear Laboratory, Durham, 
             North Carolina 27708, USA}
\author{C.~Fransen}
\affiliation{Institut f\"ur Kernphysik, Universit\"at zu K\"oln, 
             50937 K\"oln, Germany}
\author{U.~Friman-Gayer}
\affiliation{Department of Physics, Duke University, Durham,
             North Carolina 27708, USA}
\affiliation{Triangle Universities Nuclear Laboratory, Durham, 
             North Carolina 27708, USA}
\author{R.~Gonzalez}
\affiliation{Triangle Universities Nuclear Laboratory, Durham, 
             North Carolina 27708, USA}
\affiliation{Department of Physics and Astronomy, University of North Carolina
             at Chapel Hill, Chapel Hill, North Carolina 27599, USA}
\author{E.~Hoemann}
\affiliation{Institut f\"ur Kernphysik, Universit\"at zu K\"oln, 
             50937 K\"oln, Germany}
\author{R.~V.~F.~Janssens}
\affiliation{Triangle Universities Nuclear Laboratory, Durham, 
             North Carolina 27708, USA}
\affiliation{Department of Physics and Astronomy, University of North Carolina
             at Chapel Hill, Chapel Hill, North Carolina 27599, USA}
\author{S.~R.~Johnson}
\affiliation{Triangle Universities Nuclear Laboratory, Durham, 
             North Carolina 27708, USA}
\affiliation{Department of Physics and Astronomy, University of North Carolina
             at Chapel Hill, Chapel Hill, North Carolina 27599, USA}
\author{M.~D.~Jones}
\affiliation{Triangle Universities Nuclear Laboratory, Durham, 
             North Carolina 27708, USA}
\affiliation{Department of Physics and Astronomy, University of North Carolina
             at Chapel Hill, Chapel Hill, North Carolina 27599, USA}
\author{J.~Kleemann}
\affiliation{Technische Universit\"at Darmstadt, Department of Physics, Institute for Nuclear Physics,
            64289 Darmstadt, Germany}
\author{Krishichayan}
\affiliation{Department of Physics, Duke University, Durham,
             North Carolina 27708, USA}
\affiliation{Triangle Universities Nuclear Laboratory, Durham, 
             North Carolina 27708, USA}
\author{D.~R.~Little}
\affiliation{Triangle Universities Nuclear Laboratory, Durham, 
             North Carolina 27708, USA}
\affiliation{Department of Physics and Astronomy, University of North Carolina
             at Chapel Hill, Chapel Hill, North Carolina 27599, USA}
\author{D.~O'Donnell}
\affiliation{CEPS, University of the West of Scotland, Paisley PA1 2BE,
             United Kingdom}
\affiliation{SUPA, Scottish Universities Physics Alliance, United Kingdom}
\author{O.~Papst}
\affiliation{Technische Universit\"at Darmstadt, Department of Physics, Institute for Nuclear Physics, 64289 Darmstadt, Germany}
\author{N.~Pietralla}
\affiliation{Technische Universit\"at Darmstadt, Department of Physics, Institute for Nuclear Physics,
            64289 Darmstadt, Germany}
\author{J.~Sinclair}
\affiliation{CEPS, University of the West of Scotland, Paisley PA1 2BE,
             United Kingdom}
\affiliation{SUPA, Scottish Universities Physics Alliance, United Kingdom}
\author{V.~Werner}
\affiliation{Technische Universit\"at Darmstadt, Department of Physics, Institute for Nuclear Physics,
            64289 Darmstadt, Germany}
\author{O.~Wieland}
\affiliation{INFN, Sezione di Milano, 20133 Milano, Italy}
\author{J.~Wilhelmy}
\affiliation{Institut f\"ur Kernphysik, Universit\"at zu K\"oln, 
             50937 K\"oln, Germany}



\date{\today}

\begin{abstract}

\begin{description}
\item[Background] Photon strength functions are an important
ingredient in calculations relevant for the nucleosynthesis of heavy
elements. The relation to the photoabsorption cross section allows to
experimentally constrain photon strength functions by investigating the
photo-response of atomic nuclei.
\item[Purpose] We determine the photoresponse of $^{66}$Zn in the
energy region of 5.6 MeV to 9.9 MeV and analyze the contribution of the
'elastic' decay channel back to the ground state. In addition, for the elastic
channel electric and magnetic dipole transitions were separated.
\item[Methods] Nuclear resonance fluorescence experiments were
performed using a linearly-polarized quasi-monoenergetic photon beam
at the High Intensity $\gamma$-ray Source. Photon beam energies from 5.6 to 9.9 MeV with an energy spread of about
3\% were selected in steps of 200-300 keV. Two High Purity Germanium
detectors were used for the subsequent $\gamma$-ray spectroscopy.
\item[Results] Full photoabsorption cross sections are extracted
from the data making use of the monoenergetic character of the photon
beam. For the ground-state decay channel, the average contribution of electric and
magnetic dipole strengths is disentangled. 
The average branching ratio back to the ground state is determined as well.
\item[Conclusions] The new results indicate lower cross sections when compared to the values extracted from a former experiment using bremsstrahlung on $^{66}$Zn.
In the latter, the average branching ratio to the ground state is 
estimated from statistical-model calculations in order to analyze the
data. 
Corresponding estimates from statistical-model calculations underestimate this branching ratio compared to the values extracted from the present analysis, which would partly explain the high cross sections determined from the bremsstrahlung data.
\end{description}

\end{abstract}


\maketitle


\section{Introduction}

The modeling of the nucleosynthesis of heavy elements depends on input
from nuclear structure and nuclear reaction studies. 
Statistical model calculations within the Hauser-Feshbach formalism~\cite{haus52} are needed for estimating the relevant reaction rates. 
One of the important quantities usually used for such
calculations is photon strength functions (PSF). 
These characterize the average probability for the
emission and absorption of photons by atomic nuclei. Experimentally,
PSF are studied using a variety of reactions, see Refs.~\cite{gori19, lars19} for an
overview. Among the methods, photon-induced reactions offer the
possibility to study PSF via their connection to the photoabsorption
cross sections $\sigma_{\gamma}$.  

Photonuclear reactions are a common
tool to study a variety of nuclear structure phenomena
~\cite{zilg22}. Due to the well-understood interaction, properties of
excited nuclear states can be determined in a model independent
way. Below the particle-emission threshold, nuclear resonance fluorescence (NRF)
experiments are one of the work horses to provide a broad
experimental database.

An interesting region in excitation energy to study the photoresponse of nuclei is from 5 to 10 MeV. In this energy region, the so-called Pygmy Dipole
Resonance (PDR) is found in a number of nuclei \cite{savr13, brac19}. In the
medium mass range, a number of NRF experiments have been performed in
order to study the PDR, e.g., in the Z=28 \cite{bauw00,sche13, sche13a}
isotopes or the N=50 \cite{ruse09,schw13,maki16, wilh20} isotones. 
Recent findings on the neutron-number dependence of the low-energy dipole strength near the shell closure $N=28$~\cite{pai13, kris15, shiz17, wilh18, ries19} indicate its sensitivity to the valence shell and, thus, to single-particle effects~\cite{ries19}.
In particular, the Ni and Zn isotopic chains offer the possibility to
study the PDR over a broad range of neutron-to-proton ratios in medium
mass nuclei.

In order to study the photoresponse of $^{66}$Zn in the energy region
up to its neutron separation energy ($S_n = 11.06$ MeV), nuclear
resonance fluorescence (NRF) experiments were performed at the
$\gamma$ELBE facility~\cite{schw05} using broad-band bremsstrahlung as
well as at the High Intensity $\gamma$-ray Source (HI$\gamma$S)
facility~\cite{well09} making use of a horizontally linearly-polarized, 
quasi-monoenergetic photon beam produced via
Laser-Compton-Backscattering (LCB). In the analysis presented in
Ref.~\cite{schw21}, the excitation strength of isolated resonances as well
as the photoabsorption cross section in bins of 100 keV were
extracted from the data taken with bremsstrahlung. The method to
determine the latter requires an accurate description of the decay
behavior of the photo-excited states within statistical model
calculations as described, e.g., in Refs.~\cite{schw07, ruse08, mass12, maki16, schw20, mues20, wilh20}.  The parity-quantum numbers of isolated
excited states were determined exploiting the linear polarization of
the LCB photon beam produced at HI$\gamma$S. Furthermore, as shown for
other cases~\cite{tonc10, isaa13, romi13, godd13, isaa19, isaa21}, the
monoenergetic character of the LCB photon beam allows to determine
the full photoabsorption cross section averaged over the energy
profile of the photon beam. At the same time, the average branching
ratio needed in the analysis of the bremsstrahlung data can be
extracted from the LCB data as well, which allows to verify the
accuracy of the calculations within the statistical model. In the
present manuscript, we show the analysis of the LCB data and compare
the results with the data presented in \cite{schw21}.

\section{Experiment}

An extensive description of the setup and relevant details of the
experiments are already given in \cite{schw21}. Therefore, only a brief
summary is provided here. The experiment was performed at the
HI$\gamma$S facility of the Triangle Universities Nuclear Laboratory at Duke University in Durham, NC, USA
\cite{well09}. At the time of the experiment, two NRF setups were
available, the $\gamma^{3}$ setup \cite{loeh13} and a second one
located downstream, which was used for the measurement on
$^{66}$Zn. Two large-volume High Purity Germanium (HPGe) detectors
were positioned at polar angles of $90^{\circ}$ with respect to the incoming LCB beam. 
One HPGe of 100\% relative efficiency was placed in the polarization plane (hereafter referred to as \textit{horizontal} detector) and another one of 80\% relative efficiency was located perpendicular to the polarization plane (hereafter referred to as \textit{vertical} detector). This geometry is ideal to separate electric dipole ($E1$) from magnetic dipole ($M1$) transitions in an even-even nucleus
\cite{piet02b}. The target consisted of 1.4993~g of zinc enriched to
98\% in $^{66}$Zn, and was formed to a disc with a diameter of 20~mm,
which is about the size of the photon beam. For the present analysis,
data taken at beam energies of 5.6, 5.75, 5.9, 6.1, 6.3, 6.5, 6.7,
6.9, 7.1, 7.3, 7.5, 7.7, 7.9, 8.15, 8.4, 8.65, 8.9, 9.15, 9.4, 9.65,
and 9.9~MeV are considered. For the determination of full energy peak
detection efficiencies, a set of calibration measurements have been
taken before and after the experiments using standard radioactive
sources such as $^{56}$Co, $^{60}$Co, and $^{152}$Eu.

\section{Analysis}

The present analysis focuses on the extraction of the photoabsorption
cross section averaged over the beam energy profiles of the single
measurements. For this purpose, we follow mostly the formalism outlined in
\cite{isaa21}.

The photoabsorption cross section $\sigma_{\gamma}$ can be expressed
by the sum of the so-called ``elastic'' part $\sigma_{\gamma \gamma}$
and the ``inelastic'' $\sigma_{\gamma \gamma '}$
\begin{align}\label{equ::cs}
\sigma_{\gamma} &= \sigma_{\gamma \gamma} + \sigma_{\gamma \gamma '} \nonumber \\
&= \frac{Y_{0 \rightarrow x \rightarrow 0}}{N_{T}N_{\gamma }} + \frac{Y_{0
\rightarrow x \rightarrow i}}{N_{T}N_{\gamma}} 
\end{align}
with $N_{T}$ and $N_{\gamma}$ being the number of target nuclei and
integrated number of impinging photons, respectively.  The first part
$Y_{0 \rightarrow x \rightarrow 0}$ is the sum over the reaction
yields of all photo-excited states $x$ decaying back to the ground
state (elastic channel), while $Y_{0\rightarrow x \rightarrow i}$ represents the sum of
all reactions with decays via lower-lying excited states (inelastic channel). The
monoenergetic character of the LCB photon beam allows to separately extract both
quantities from the measured spectra. The number of
target nuclei is calculated from the mass of the NRF target, and
the photon intensity can be calibrated via known excitation cross
sections of isolated single states determined in the experiment using
bremsstrahlung \cite{schw21}.

\subsection{Photon-flux}\label{sub::flux}

In NRF experiments with bremsstrahlung, absolute excitation cross
sections for single excitations can be determined relative to a
well-known calibration standard, usually $^{11}$B or $^{27}$Al. For
experiments using quasi-monoenergetic LCB photon beams, this is not
possible, since these standards do not have excited states in all
the required excitation-energy regions. Moreover, the excitations from
the calibration standard might overlap with transitions from 
the nucleus of interest in the measured $\gamma$-ray spectra. 
However, since energy-integrated cross sections for isolated states 
of $^{66}$Zn have been already determined relative to the calibration standard $^{11}$B from NRF measurements with bremsstrahlung~\cite{schw21}, these known transitions
can be used to calibrate the photon flux in the present case. This is one of the
reasons combined experiments using bremsstrahlung and LCB photon
beams are so powerful.

\begin{figure}
 \includegraphics[width=\columnwidth]{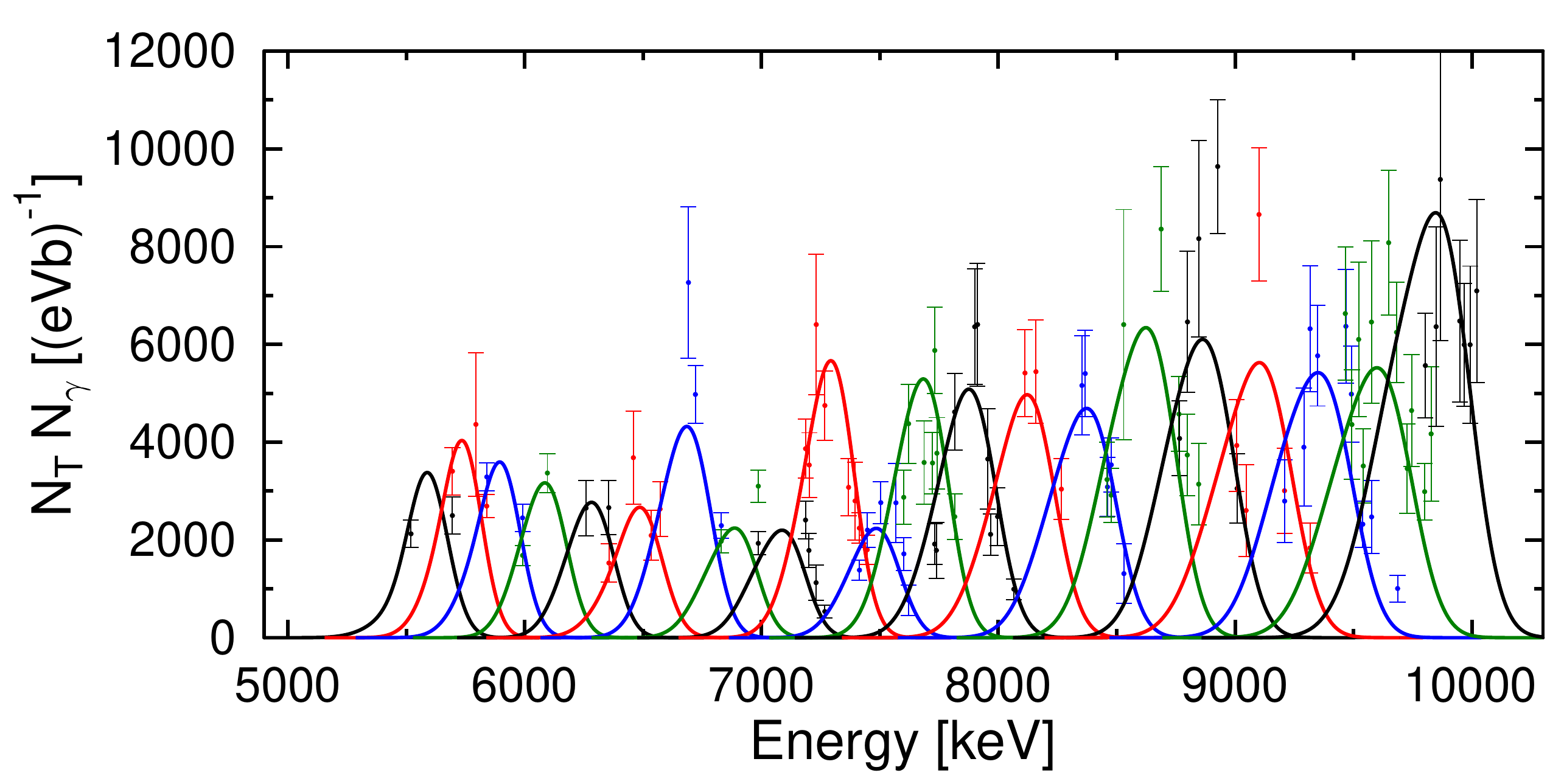}%
 \caption{\label{fig::ng} (Color online) Normalized LCB beam spectral distributions fit to values extracted from known individual excited states. The different colors are meant to improve visual distinction of the individual data sets and respective beam profiles. See text for details.}
\end{figure}

In our analysis, we applied the method first presented in
\cite{tamk19} and later applied in Ref.~\cite{paps20} as well. Instead
of normalizing the measurements for each beam energy separately (as
done e.g. in \cite{isaa21}), in a first step the single measurements
were normalized to each other and then normalized to all observed
states at all energies with one single parameter. The first step can
be done by making use of the fact, that the low-energy part of the
spectra originates from (atomic) reactions of the photon beam in the
target and, thus, is directly proportional to the integrated
photon-beam intensity. For the relative normalization, the full
geometry was implemented in the $utr$ simulation tool kit~\cite{utr19}
based on \texttt{GEANT4}~\cite{agos03, alli06, alli16}, and a photon
beam with the corresponding energy impinging on the target was
simulated. For each beam energy, the simulated spectrum was fitted to
the low-energy part of the measured spectrum by the intensity in the
511-keV annihilation peak.  It should be noted that this is a very
similar approach to that described in Refs.~\cite{tamk19, paps20},
with the difference that a well-defined peak is used for normalization
instead of a continuous energy region slightly above 511 keV.  The
extracted factors are proportional to the integrated photon flux in
each measurement and were used to normalize the photon flux
distributions, which were measured at the beginning of the experiment
for each beam-energy setting using an in-beam HPGe detector. After
this procedure, the photon-flux distributions, which were normalized
relative to each other, differ by only one global scaling factor from
the absolute photon-flux intensity. This global factor was determined
in the last step by a simultaneous fit of all distributions to the
energy-integrated cross section values extracted for known single
excitations observed in the spectra. The result is shown in
Fig.~\ref{fig::ng}. The values for the integrated photon flux
$N_{\gamma}$ are calculated by integrating the single photon flux
distributions.

The advantage of this procedure is a reduced statistical uncertainty
for energy settings, for which only a few known excitations can be
used as calibration points shown in Fig.~\ref{fig::ng}. The pure
statistical (internal) uncertainty of the global scaling factor in the
present case is 1.9\%, while the weighted (external) uncertainty is 3.1\%,
which accounts for the spread of the data about the fit. The
latter we used as the uncertainty in the calculation of the cross
sections below.

\subsection{Elastic cross section}\label{sub::elastic}

\begin{figure}
 \includegraphics[width=\columnwidth]{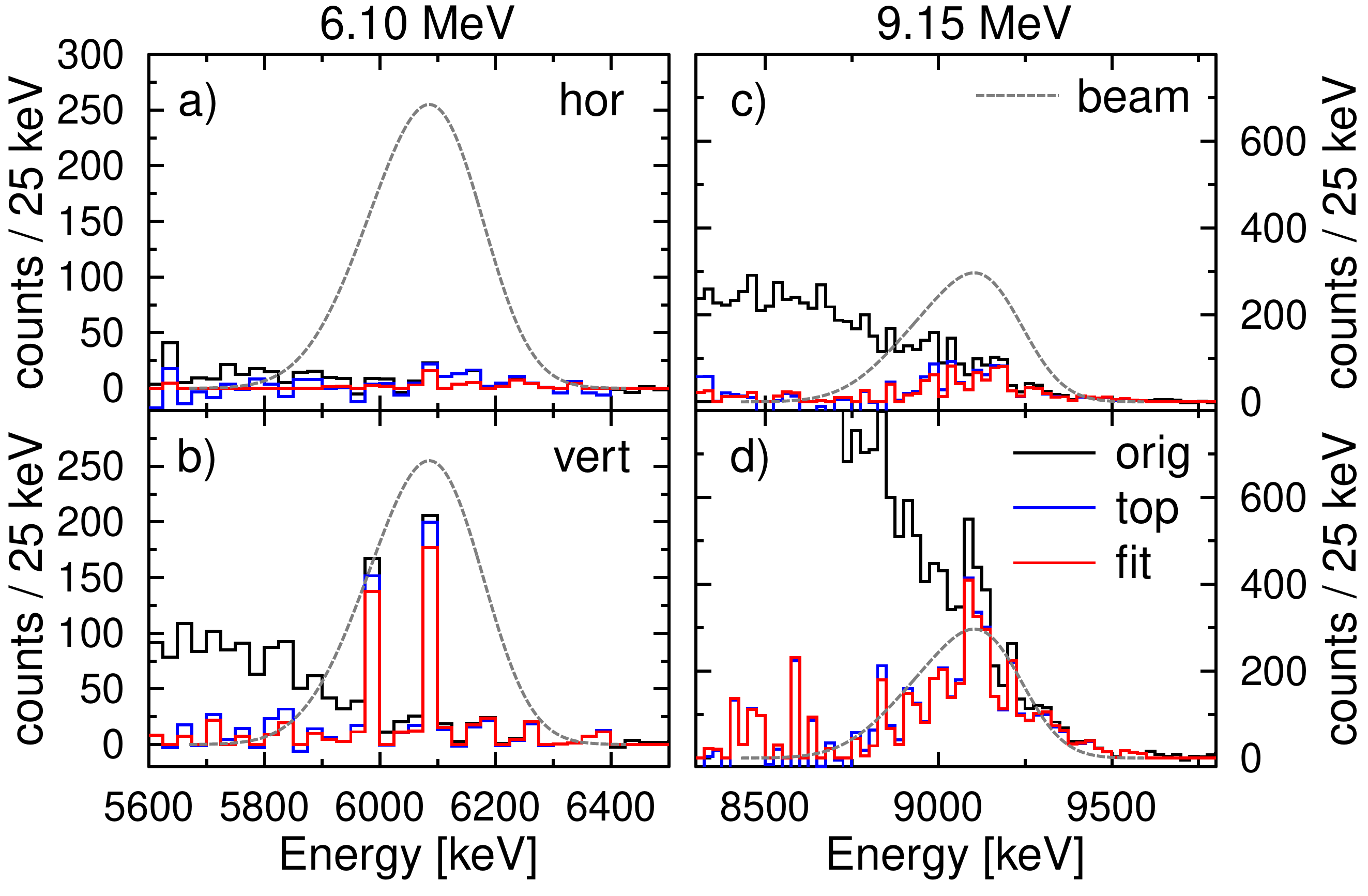}%
\caption{\label{fig::spectra} (Color online) Spectra recorded with the horizontal (upper panels) and vertical (lower panels) detector, respectively, in 25-keV binning for a LCB beam energy of 6.1~MeV (left) and 9.15~MeV (right). Original spectra are displayed in black, deconvoluted spectra in blue (top-down method) and red (fitting method). In addition, the energy profile of the photon beam is given (dashed line).}
\end{figure}

The state-by-state analysis of isolated resonances does have a
sensitivity limit that depends on the background in the $\gamma$-ray
spectra; i.e., a particular nuclear level needs to have a minimum NRF scattering 
cross section in order to be observable. Thus, the sum of all observed 
individual cross sections does not add up to
the average elastic cross section $\sigma_{\gamma \gamma}$. In
comparison to microscopic models, this can be accounted for by analyzing the fragmentation of the observed dipole strength of individual excitations
as done in Ref.~\cite{savr08a}. However, a correction of the summed experimental cross section based
on this comparison would include a model dependency. The NRF method
using an LCB photon beam allows instead a clean determination of the average
elastic cross section: Since, in the excitation process, the photon is
always fully absorbed, the intensity of the measured $\gamma$ rays 
emitted in the subsequent ground-state transitions observed at energies 
corresponding to the incident LCB beam is directly proportional to the number of elastic NRF
reactions $Y_{0 \rightarrow x \rightarrow 0}$
\begin{align}\label{equ::elastic}
A_{0 \rightarrow x \rightarrow 0} =  {Y_{0 \rightarrow x \rightarrow 0}} \cdot {\int\limits_{\Delta \Omega} \epsilon(E_{x},\Omega) \cdot W_{0 \rightarrow x \rightarrow 0} \cdot d\Omega}
\end{align}
with $A_{0 \rightarrow x \rightarrow 0}$ being the integrated
intensity of the spectrum at $E_{x}$ after correcting for the detector
response. The integral in Eq.~(\ref{equ::elastic}) includes the angular distribution $W_{0 \rightarrow x \rightarrow 0}$ of the NRF reaction and the energy and angle-dependent efficiency $\epsilon(E_{x},\Omega)$, which is simulated using \texttt{GEANT4}
as described above. The same simulation is used to determine the
energy-dependent detector response, which is used to deconvolute the
measured $\gamma$-ray spectra, and extract the original intensity. In
the present analysis, we used two different deconvolution methods, the
top-down method (see, e.g., \cite{mass12}), and the fitting procedure described in Refs.~\cite{loeh16, isaa16b, horst}. Figure~\ref{fig::spectra} shows the original spectra for two beam energies
for both detectors together with the results of the
deconvolution. Parameters within the deconvolution procedures are the
used bin size, the low and high energy cut-off and the detector
response itself. The parameters were varied as well as
the geometry used for the response simulation within reasonable values
in order to calculate the systematic uncertainties, which are given in
the next section together with the results.

As can be seen in Fig.~\ref{fig::spectra}, the intensity in the
horizontal detector is much smaller compared to that in the vertical detector
(both have about the same efficiency), which indicates that $E1$ is the
dominant multipolarity of the elastic photoresponse of $^{66}$Zn. However, making use of the angular
distribution, the relative contribution of $E1$ and $M1$ can be
disentangled by rewriting Eq.~(\ref{equ::elastic}) as
\begin{align}\label{equ::elastic2}
A^{i}_{0 \rightarrow x \rightarrow 0} = ~ &Y_{0 \rightarrow 1^{-} \rightarrow 0} \int\limits_{\Delta \Omega} \epsilon^{i}(E_{x},\Omega) W^{i}_{0 \rightarrow 1^{-} \rightarrow 0}  d\Omega \nonumber \\
+ ~ &Y_{0 \rightarrow 1^{+} \rightarrow 0}  \int\limits_{\Delta \Omega} \epsilon^{i}(E_{x},\Omega)  W^{i}_{0 \rightarrow 1^{+} \rightarrow 0}  d\Omega
\end{align}
where the index $i$ stands for the two detectors. Thus, measuring the
intensities $A^{i}$ at two different angles is sufficient to separate
$E1$ and $M1$ reaction yields; see also Ref.~\cite{isaa21}. For the given
geometry, $E2$ and $M1$ transitions have the same angular distribution and, thus,
cannot be distinguished. However, a considerable contribution of $E2$ at
the given excitation energies seems very unlikely and we, thus,
neglect this in the following.

\subsection{Inelastic cross section}\label{sub::inelastic}

\begin{figure}
 \includegraphics[width=0.8\columnwidth]{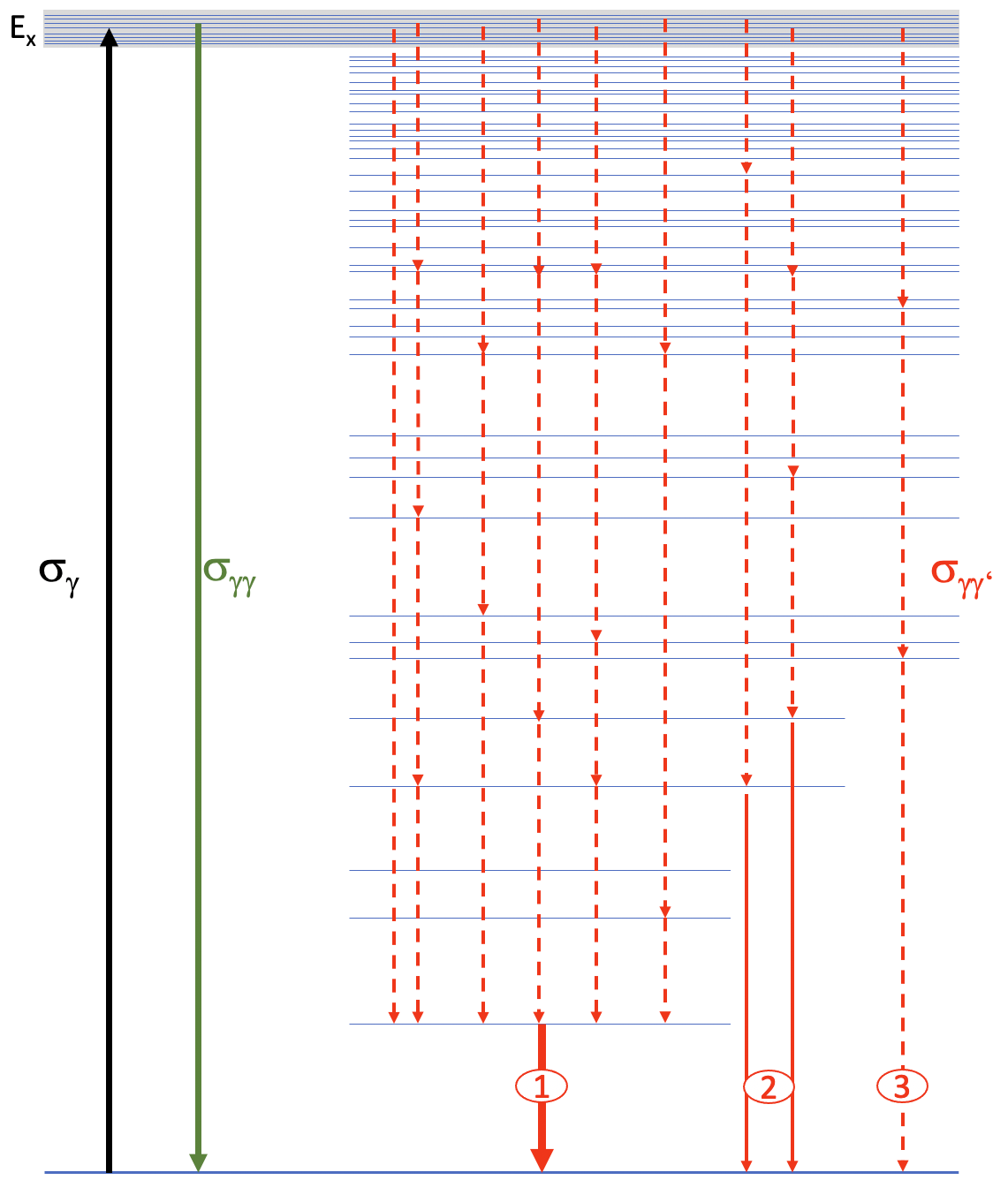}%
\caption{\label{fig::levels} (Color online) Illustration of the
different decay paths and contributions to the total photoabsorption
cross sections $\sigma_{\gamma}$. After exciting the nucleus to the
energy region around $E_{x}$ defined by the energy profile of the
photon beam the primary excited levels can either elastically decay back
to the ground state ($\sigma_{\gamma \gamma}$ in green) or inelastically via
intermediate states ($\sigma_{\gamma \gamma '}$ in red). Most of the
latter intensity is collected by the first excited $2^{+}$ state (case
1). At higher excitation energies ground-state transitions of higher-lying states become visible (case 2), which are added to the
$\sigma_{\gamma \gamma '}$ contributions. Possible transition
bypassing completely the first excited states stay unobserved (case
3).  }
\end{figure}

The determination of the inelastic reaction yield $Y_{0 \rightarrow x
\rightarrow i}$ from the primary decay of the initially excited states
is difficult, since in most cases the average decay branchings
are small \cite{loeh16,isaa19}. However, most of the decay
cascades will, at some point, reach the first or one of the other low
excited states. Schematically, this is illustrated in
Fig.~\ref{fig::levels}: The narrow bandwidth ensures an excitation
into a small excitation energy region at $E_{x}$, which enables a
determination of the full elastic contribution as outlined
above. Decay paths going via intermediate excited states will end up
in most cases in one of the lowest excited states. Therefore, the
intensity of the decay of these excited states is a good
approximation of the number of inelastic reactions.

Using this method to measure $\sigma_{\gamma \gamma '}$ has first been
proposed by Tonchev et al. \cite{tonc10}, and was applied in a number
of cases, see e.g. Refs.~\cite{isaa13, romi13, godd13, sche13, wilh20,
isaa21}.

\begin{figure}
 \includegraphics[width=\columnwidth]{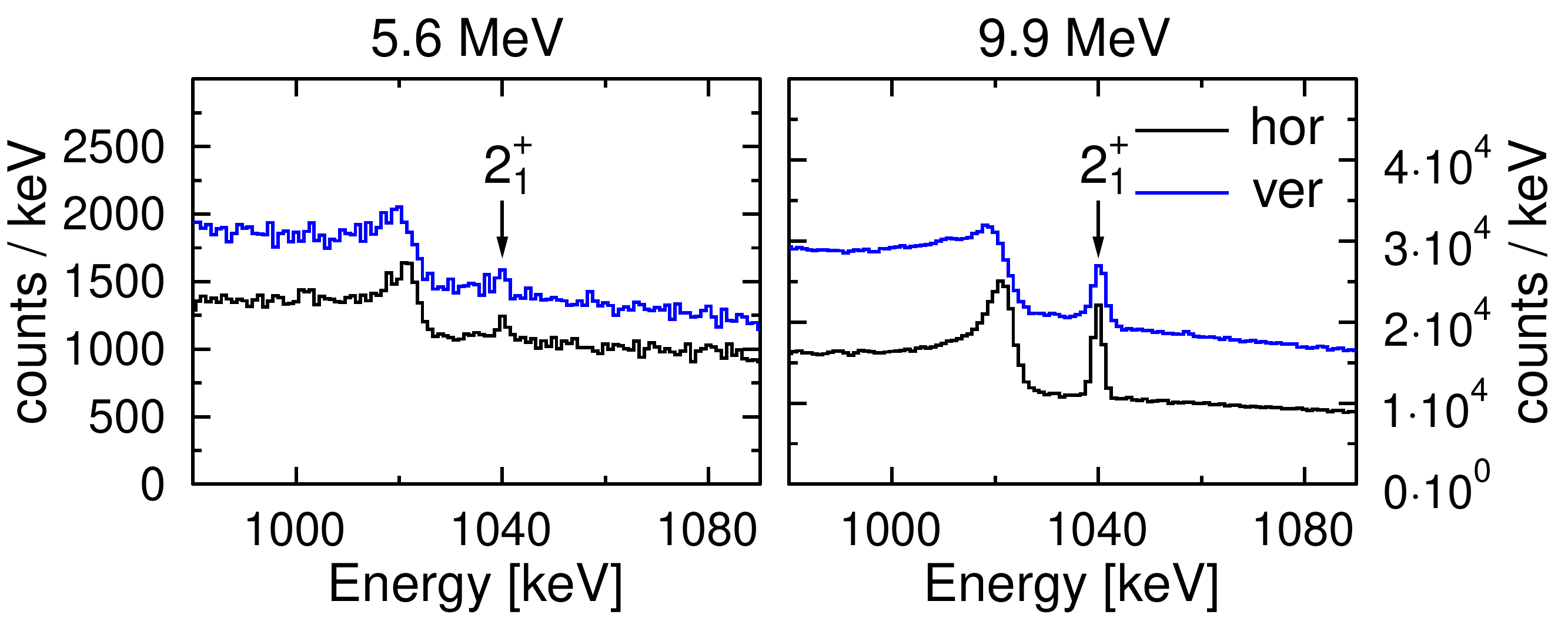}%
 \caption{\label{fig::spec-low} (Color online) Measured spectra in the region of the
decay energy of the first excited $2_{1}^{+}$ state at 1039 keV
(marked with an arrow). At the lowest beam energy of 5.6~MeV, the peak is hardly
visible above the background, while at the highest energy of 9.9~MeV a strong
feeding of the state is observed. The structure left to the peak is
background resulting from the sum of two 511-keV annihilation $\gamma$ rays.}
\end{figure}

In the present case of $^{66}$Zn, the decay of the first excited
$2_{1}^{+}$ state at 1039 keV is hardly visible above the background
at the lowest beam energies, see Fig.~\ref{fig::spec-low}. With
increasing beam energy, the feeding of the $2_{1}^{+}$ state becomes
increasingly pronounced and, thus, the inelastic part of the
photoabsorption cross section more important. At energies above 7.5
MeV, two additional peaks appear, that originate from decays of
higher-lying states which do not entirely decay via the $2_1^+$ state,
namely the $2_3^+$ at 2780.2 keV and the $1_1^-$ state at 3380.9 keV
(case 2 in Fig.~\ref{fig::levels}). The intensities of the
ground-state decays of these two states are also included in the
determination of $\sigma_{\gamma\gamma^\prime}$, but contribute only
about 10\% even at the highest beam energies. This demonstrates, that
cascades completely bypassing the first excited states (case 3 in
Fig.~\ref{fig::levels})) will only have a small impact on the measured
inelastic cross section. This is the only contribution of
$\sigma_{\gamma\gamma^\prime}$ missed in the present analysis.

\section{Results}

In order to determine the integrated intensities $A^{i}_{0 \rightarrow
x \rightarrow 0}$ for the two detectors positioned vertically ($i=\mathrm{ver}$)
and horizontally ($i=\mathrm{hor}$) with respect to the polarization plane, the
corresponding unfolded spectra were integrated in the region where the
beam intensity is above 10\% of its maximum. This region has been
varied together with parameters within the unfolding procedure (as
mentioned before) in order to derive systematic errors for the
extracted cross sections. By this integration, resolved peaks as well as so-called
unresolved contributions are included.

\begin{figure}
 \includegraphics[width=\columnwidth]{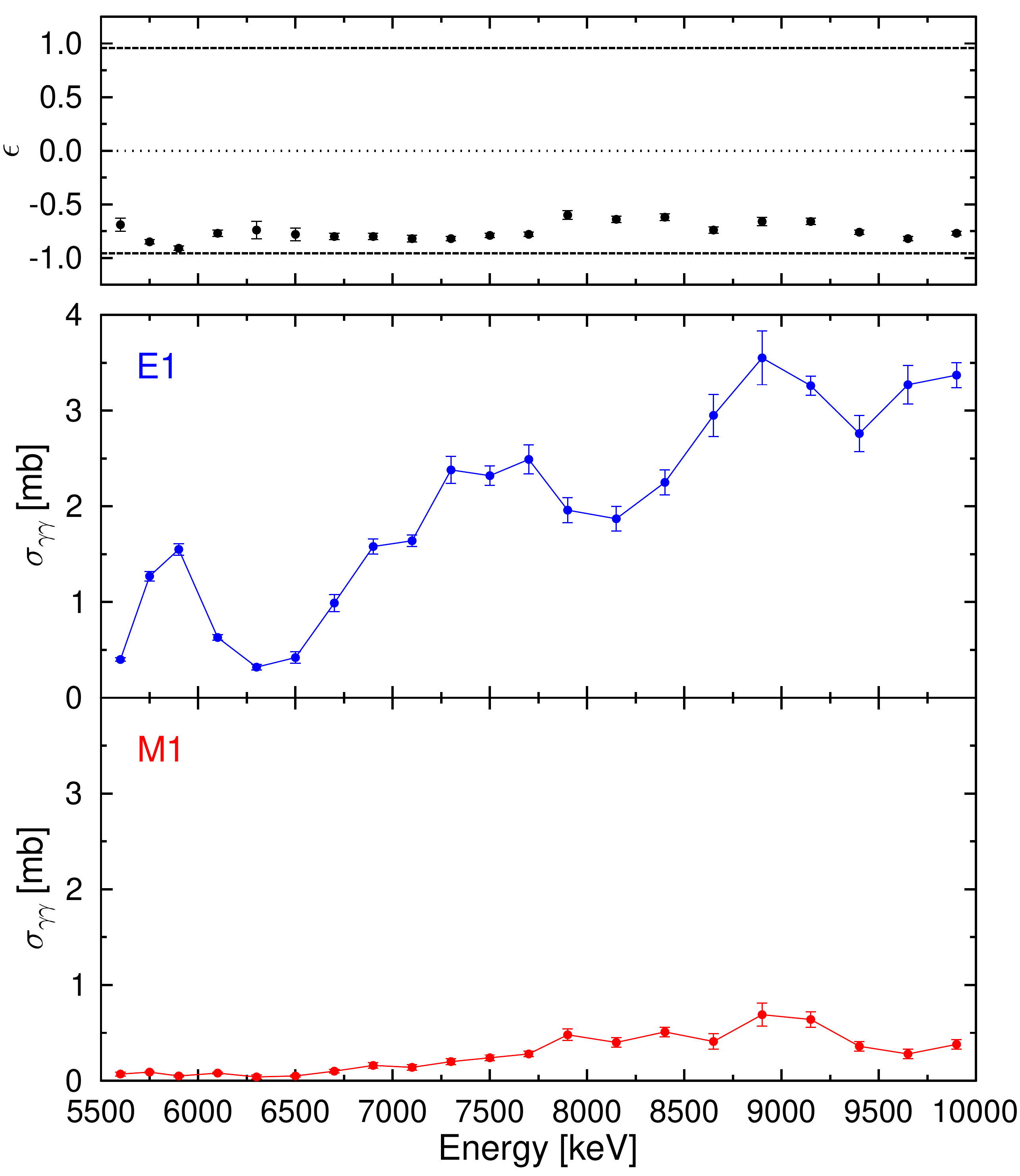}%
 \caption{\label{fig::elastic} (Color online) Upper panel: Average asymmetry determined from Eq.~(\ref{equ::asym}). Lower panel: Elastic $E1$ and $M1$ cross section
 separated by analyzing the angular distribution of the emitted ground-state $\gamma$-ray decays. Clearly, $E1$ transitions dominate over the energy region under investigation. The numerical data displayed can be found in Ref.~\cite{tudatalib}.}
\end{figure}

The average asymmetry $\epsilon$ is defined by
\begin{align}\label{equ::asym}
  \epsilon = \frac{A^{\mathrm{hor}}_{0 \rightarrow x \rightarrow 0}-A^{\mathrm{ver}}_{0
      \rightarrow x \rightarrow 0}}{A^{\mathrm{hor}}_{0 \rightarrow x \rightarrow
  0}+A^{\mathrm{ver}}_{0 \rightarrow x \rightarrow 0}}
\end{align}
and is presented in the upper panel of Fig.~\ref{fig::elastic}, together with
the expected values of the angular distributions for $E1$ and $M1$ ground-state transitions, respectively. 
The latter are determined via simulations and are also used
in Eq.~(\ref{equ::elastic}). 
The results for the elastic cross section separated by $E1$ and $M1$
contribution are given in the lower panel of Fig.~\ref{fig::elastic}. 
As mentioned before, the $E1$ cross section
is dominant at all excitation energies. The separation into $E1$ and $M1$
contributions is only possible for the elastic part of the cross
section.
With the method to extract the inelastic cross section one cannot discriminate between the two contributionscontributions and there is no reason to assume
that the branching of the initially populated $J^{\pi}=1^{-}$ and $J^{\pi}=1^{+}$
states is the same.

\begin{figure}
 \includegraphics[width=\columnwidth]{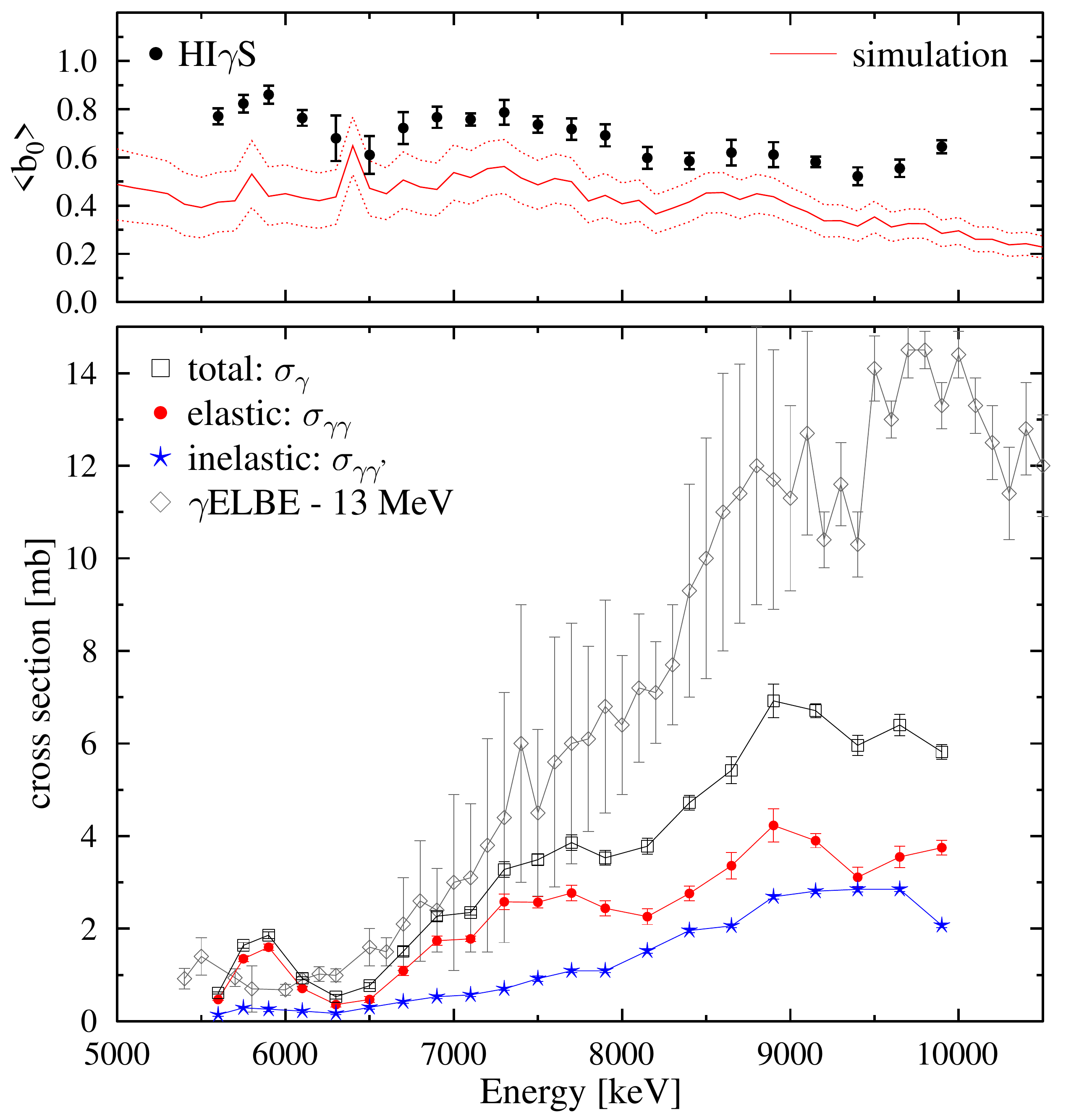} \caption{\label{fig::cs}
(Color online) Lower panel: Elastic, inelastic and total cross
sections determined within the present analysis (numerical values can be found in Ref.~\cite{tudatalib}) compared to the
results obtained from bremsstrahlung data taken at $\gamma$ELBE
\cite{schw21}. For our new results systematic uncertainties are
plotted. Upper panel: Mean branching ratios $\langle b_0 \rangle$
determined from the present results (black data points) and extracted
from statistical model calculations (red solid line) with its $1
\sigma$ uncertainty band (red dotted lines) taken from \cite{schw21}.}
\end{figure}

The results for the elastic, inelastic and total cross sections are
shown in the lower panel of Fig.~\ref{fig::cs}. At low excitation energy, $\sigma_{\gamma
\gamma '}$ hardly contributes to the total photoabsorption cross section, 
while it continuously increases towards 10 MeV. 
As in other cases, the inelastic cross section exhibits fewer
structures when compared to the elastic cross section, which is probably 
mostly driven by strong single excitations. The present results are
compared to those from Ref.~\cite{schw21} which have been obtained by
using the data from NRF experiments with bremsstrahlung. 
It is obvious that, towards
10 MeV, there is an increasing discrepancy between the two data sets. At
10 MeV, the present value for the total photoabsorption cross section is only about
half of that reported in \cite{schw21}.

An important quantity needed within the analysis in \cite{schw21} is
the decay pattern of the excited states involved. In \cite{schw21}, this 
is modeled by simulations of statistical $\gamma$-ray cascades using
the code $\gamma$DEX \cite{mass12,schr12}. With the present data, the
average ground-state branching ratio $\langle b_{0} \rangle =
\sigma_{\gamma\gamma} / \sigma_{\gamma}$ can be 
determined experimentally as a function of the excitation energy. The upper panel of
Fig.~\ref{fig::cs} compares the data to the corresponding values
of the cascade simulations used in \cite{schw21}. While the shapes
roughly agree, the absolute simulated values amount on average to
about 60\% of the experimental ones. This difference explains a large fraction of
the discrepancy between the cross sections extracted in the present
analysis and in the statistical analysis of the bremsstrahlung data
taken at $\gamma$ELBE. Moreover, this demonstrates the importance of the
determination of model-independent branching ratios and, hence,
photoabsorption cross sections below the neutron-separation energy.
However, the region above 10 MeV was not covered in the
experiment at HI$\gamma $S.
The experimental results for $\sigma_{\gamma}$, $\braket{b_0}$, and the ratio of the $M1$ to $E1$ elastic cross sections are summarized in Table~\ref{tab::cs}.

\begin{table}
\caption{\label{tab::cs} Total photoabsorption cross section
$\sigma_{\gamma}$, average ground-state branching ratio $\langle b_0 \rangle$ and the ratio of
$M1$ to $E1$ elastic cross section deduced within the present
analysis. For $\sigma_{\gamma}$, statistical and systematic uncertainties are
given separately.}
\begin{ruledtabular}
\begin{tabular}{cccccc}
$E_{x}$[MeV] & $\sigma_{\gamma }$[mb] ($\pm$  stat.) ($\pm$  sys.) & $\langle b_0 \rangle$ & $M1$/$E1$ \\[0cm]
\hline
      5.60  &   0.61(5)(2)   &  0.74(4)  &     0.18(4)  \\[0cm]
      5.75  &   1.64(6)(6)   &  0.81(3)  &     0.07(1)  \\[0cm]
      5.90  &   1.86(9)(7)   &  0.86(3)  &     0.03(1)  \\[0cm]
      6.10  &   0.93(5)(3)   &  0.74(4)  &     0.12(2)  \\[0cm]
      6.30  &   0.53(4)(5)   &  0.65(6)  &     0.14(6)  \\[0cm]
      6.50  &   0.77(6)(6)   &  0.58(8)  &     0.11(4)  \\[0cm]
      6.70  &   1.51(7)(10)  &  0.70(6)  &     0.10(2)  \\[0cm]
      6.90  &   2.27(10)(10) &  0.75(4)  &     0.10(2)  \\[0cm]
      7.10  &   2.35(11)(6)  &  0.74(3)  &     0.09(2)  \\[0cm]
      7.30  &   3.28(11)(17) &  0.77(5)  &     0.08(2)  \\[0cm]
      7.50  &   3.49(15)(12) &  0.72(3)  &     0.11(1)  \\[0cm]
      7.70  &   3.86(13)(17) &  0.70(4)  &     0.11(1)  \\[0cm]
      7.90  &   3.53(12)(16) &  0.64(4)  &     0.24(4)  \\[0cm]
      8.15  &   3.78(13)(17) &  0.55(4)  &     0.21(3)  \\[0cm]
      8.40  &   4.72(17)(16) &  0.53(3)  &     0.23(3)  \\[0cm]
      8.65  &   5.42(18)(29) &  0.59(4)  &     0.14(3)  \\[0cm]
      8.90  &   6.92(22)(36) &  0.57(4)  &     0.19(4)  \\[0cm]
      9.15  &   6.71(22)(15) &  0.54(2)  &     0.20(2)  \\[0cm]
      9.40  &   5.96(20)(22) &  0.49(3)  &     0.13(2)  \\[0cm]
      9.65  &   6.40(21)(23) &  0.53(3)  &     0.09(2)  \\[0cm]
      9.90  &   5.82(19)(16) &  0.62(2)  &     0.11(2)  \\[0cm]
\end{tabular}
\end{ruledtabular}
\end{table}

\section{Conclusions}

We have determined the photoabsorption cross sections of $^{66}$Zn in
the energy range between 5.6 and 10 MeV using NRF data taken with
a quasi-monoenergetic and linearly-polarized photon beam at
HI$\gamma$S. While the monoenergetic character of the beam allows to
extract the full elastic as well as inelastic cross sections, the high
degree of linear polarization enables for a separation of the $E1$ and $M1$
contributions to the elastic cross sections.

Above about 8~MeV, the determined total photoabsorption cross sections
are considerably smaller than the ones obtained from statistical-model calculations applied to the interpretation of NRF experiments using continuous-energy bremsstrahlung, only.
The discrepancy is increasing with excitation energy to about
a factor of two at 10 MeV. One of the important ingredients in the
analysis of the bremsstrahlung data is the decay behavior of the
photo-excited states, which in that case is estimated through simulations of statistical
$\gamma$-ray cascades. Within the present analysis, the average
ground-state branching ratios $\langle b_0 \rangle$ are deduced from
experimental intensities, and are larger than the
simulated ones by about 60\%. This partly explains the differences between the
present cross sections and the ones from the bremsstrahlung
experiment. However, a small amount of the intensity of
inelastic transitions bypassing the low-energy excited states states may be
missing in the present analysis.

These differences may indicate that certain inputs to the statistical
model, such as nuclear level densities, are not well described in nuclei
around $A$ = 60. For heavier nuclei, there are examples of better
agreement between simulated and experimental branching ratios and
cross sections~\cite{mass14b,mass12}, but discrepancies have also been reported before \cite{romi13, isaa13}.
Altogether, as shown here for $^{66}$Zn, combined experiments provide
the possibility to determine scattering cross sections of resolved
transitions over a wide energy range using bremsstrahlung and to
determine model-independent continuous photoabsorption cross sections using
LCB photon beams.

\section*{Data availability}

The datasets obtained in the experiments and used for analysis in the current study are available from the corresponding author on reasonable request. The generated results presented in this article are openly available and published in the TUdatalib data repository of Technische Universit{\"a}t Darmstadt~\cite{tudatalib}.

\section*{Acknowledgments}

The authors thank the HI$\gamma$S accelerator staff for providing
excellent LCB photon beams and experimental conditions.
R.~M. acknowledges support by the U.S. Department of Energy, Office of
Science, Office of Nuclear Physics under grant No. LANLEM77. The UWS
group acknowledges financial support from UK STFC (grant
No. ST/P005101/1).  J.~S. acknowledges financial support by the UK
Nuclear Data Network. T.~B. and V.~W. acknowledge support by the BMBF
grant 05P21RDEN9 and J.~W. acknowledges support by the BMBF grant
05P18PKEN9. T.~B., U.~F.-G., J.~K., and O.~P. are supported by the
Deutsche Forschungsgemeinschaft (DFG, German Research Foundation) - Project-ID
279384907 - SFB1245. J.~I., J.~K., and O.~P. are supported by the grant "Nuclear Photonics" within the LOEWE program of the State of Hesse. J.~I. and N.~P. acknowledge the support by the State of Hesse within the Research Cluster ELEMENTS (Project ID 500/10.006). 
The remaining authors are supported by the US DOE, office of science, office of nuclear physics under grants No DE-FG02-97ER41033 (TUNL) and No DE-FG02-97ER41041 (UNC).

\bibliography{bibtex.bib}
\bibliographystyle{apsrev4-2}

\end{document}